\documentclass[3p]{elsarticle}
\usepackage[english]{babel}
\usepackage{amssymb}
\usepackage{amsmath}

\begin{document}

\begin{frontmatter}

\title{New exact periodical solutions of mKP-1 equation via $\overline{\partial}$-dressing}

\author[1]{V.G. Dubrovsky\corref{cor1}}
\ead{dubrovskij@corp.nstu.ru}
\author[1]{A.V. Topovsky}
\ead{topovskij@corp.nstu.ru}
\cortext[cor1]{Corresponding author}
\address[1]{Novosibirsk State Technical University, Karl Marx prospect 20, 630072, Novosibirsk, Russia.}

\begin{abstract}
We proposed general scheme for construction of exact real periodical solutions of mKP-1 equation via Zakharov-Manakov $\overline{\partial}$-dressing method,  derived convenient determinant formula for calculation of such solutions and demonstrated how reality and boundary conditions for the field $u(x,y,t)$ can be satisfied. We calculated the new classes of exact periodical solutions of mKP-1 equation:
\begin{enumerate}
  \item the class of nonsingular one-periodic  solutions or nonlinear plane monochromatic  waves;
  \item the class of two-periodic solutions without imposition of any boundary condition;
  \item the class of two-periodic solutions  with integrable boundary condition \\ $u(x,y,t)\mid_{y=0}=0$.
\end{enumerate}
We interpreted the third class of  two-periodic solutions with integrable boundary condition obtained by the use of special nonlinear superpositions of two simple one-periodical waves as eigenmodes of oscillations of the field $u(x,y,t)$ in semi-plane $y\geq 0$, the analogs of standing waves on the string with fixed endpoints.
\end{abstract}

\end{frontmatter}

\section{Introduction}
\label{Section_1}
\setcounter{equation}{0}
The famous modified Kadomtsev-Petviashvili, or shortly mKP-equation:
\begin{equation}\label{mKP}
V_{t}+ V_{xxx}-\frac{3}{2}V^2V_x+3\sigma^2\partial_x^{-1}V_{yy}-3\sigma V_x\partial_x^{-1}V_{y}=0, \quad \sigma^2=\pm 1,
\end{equation}
as 2+1-dimensional generalization of mKdV equation, in the papers  \cite{Konopelchenko_mKp} and \cite{JimboMiwa} was at the first discovered. The mKP equation can be represented as compatibility condition, in the Lax pair form $\left[L_1,L_2\right]=0$, of the following two linear auxiliary problems~\cite{Konopelchenko&Dubrovsky}-\cite{KonopelchenkoBook2}:
\begin{equation}\label{mKP auxiltary problems}
\left\{
\begin{array}{ll}
L_1\psi=\sigma\psi_y+\psi_{xx}+V\psi_x=0, \\
L_2\psi=\psi_t+4\psi_{xxx}+6V\psi_{xx}+3\left(V_x-\sigma\partial^{-1}_x V_y+\frac{1}{2}V^2\right)\psi_x+\alpha\psi=0.
\end{array}
\right.
\end{equation}
In general case the solutions of mKP equation are complex, but this equation admits two reductions:
\begin{enumerate}
  \item to pure imaginary $V=\sigma u:=iu$, $\bar{u}=u$ field $V$ ($\sigma=i$, mKP-1 case);
  \item to pure real-valued $V=\sigma u:=u$, $\bar{u}=u$ field $V$ ($\sigma=1$, mKP-2 case).
\end{enumerate}
So it is convenient to rewrite the equation (\ref{mKPv1}) in terms of the field $u$, $V=\sigma u$, in the following form~\cite{Konopelchenko&Dubrovsky}, \cite{Konopelchenko&Dubrovsky2}, \cite{KonopelchenkoBook2}:
\begin{equation}\label{mKPv1}
u_{t}+ u_{xxx}-\sigma^2\left(\frac{3}{2}u^2u_x-3\partial_x^{-1}u_{yy}+3u_x\partial_x^{-1}u_{y}\right)=0,
\end{equation}
mKP-equation in this last form admits two reductions to pure real-valued field $u$: with $\sigma=i$  for mKP-1 and  $\sigma=1$  for mKP-2 versions correspondingly.

Adequate introduction of spectral parameter $\lambda$ for mKP equation is achieved by the following transition from wave function $\psi$ to wave function $\chi$~\cite{Konopelchenko&Dubrovsky}, \cite{Konopelchenko&Dubrovsky2}, \cite{KonopelchenkoBook2}:
\begin{equation}\label{Psi&Chi}
  \Psi(x,y,t;\lambda):=\chi(x,y,t;\lambda)\exp{\left(i\frac{x}{\lambda}+
  \frac{y}{\sigma\lambda^2}+\frac{4it}{\lambda^3}\right)}.
\end{equation}
The wave function $\chi(x,y,t;\lambda)$ defined by (\ref{Psi&Chi}) admits due to (\ref{mKP auxiltary problems}) the canonical normalization $\chi\big|_{\lambda\rightarrow\infty}\rightarrow1$.
In the papers~\cite{Konopelchenko&Dubrovsky},~\cite{Konopelchenko&Dubrovsky2}
 the Riemann-Hilbert local and nonlocal problems and also $\overline\partial$-dressing method for construction of  exact solutions with functional parameters, multi-solitons and multi-lumps solutions of mKP equation (\ref{mKPv1})  have been applied.

Powerful $\overline\partial$-dressing method of Zakharov and Manakov~\cite{KonopelchenkoBook2}-\cite{Zakharov90} can be effectively applied for the construction of exact periodical solutions  of 2+1-dimensi-\\onal integrable nonlinear equations. This was demonstrated at first for calculations of periodical solutions of Nizhnik-Veselov-Novikov equation (NVN)~\cite{Dubrovsky&Topovsky&Basalaev}-\cite{Dubrovsky&Topovsky}, two-dimensional generalizations of the Kaup-Kupershmidt (2DKK) \cite{Dubrovsky&Topovsky&Basalaev2DKK} and Sawada-Kotera (2DSK) \cite{Dubrovsky&Topovsky&Basalaev2DSK} equations, Davey-Stewardson DS-1,2 equations ~\cite{Dubrovsky&Topovsky&OstreinovDS}.
$\overline\partial$-dressing method has been applied also for construction of exact periodical solutions of 2+1-dimensional integrable nonlinear equations with integrable boundaries, this was demonstrated at first in  \cite{Dubrovsky&Topovsky&OstreinovKP} where along with simply periodical solutions the periodical solutions with  integrable boundary condition $(u_{xx}+\sigma u_{y})(x,y,t)\big|_{y=0}=0$  for KP-2 equation have been calculated.


The concept of integrable boundary conditions compatible with integrability of integrable nonlinear equations by Sklyanin~\cite{Sklyanin} was at first introduced. In subsequent papers of Habibulin et all~\cite{Habibullin}-\cite{HabibullinKDV} and
others~\cite{Vereshchagin} this concept of integrable boundaries to several types of integrable nonlinear equations has been applied: for difference equations, 1+1- and 2+1-dimensional differential and integro-differential equations; a list of several integrable boundary conditions for known 2+1-dimensional nonlinear differential equations such as KP, mKP, NVN, Ishimori, etc. has been proposed and some examples of corresponding exact solutions for these equations with integrable boundary conditions have been calculated \cite{Habibullin}-\cite{Vereshchagin}.

Integrable boundaries
have been considered also in numerous papers on dromion solutions for DS, Ishimori, NVN equation, etc. \cite{AblowitzClarkson}-\cite{KonopelchenkoBook2}.
A. S. Fokas et al have been obtained interesting results via so-called Unified Approach to Boundary Value Problems \cite{FokasBook}, there was demonstrated the applicability of this method for one-dimensional and multi-dimensional linear and nonlinear differential equations.

In the present paper we constructed via $\overline\partial$-dressing method of Zakharov and Manakov new classes of exact one-periodical  and two-periodical solutions of mKP-1 equation (\ref{mKPv1}) without boundary condition and new class of exact two-periodical solutions of mKP-1 equation (\ref{mKPv1}) with integrable boundary condition ~\cite{Habibullin2}:
\begin{equation}\label{BoundaryCond}
u\big|_{y=0}=0,
\end{equation}
which is compatible with integrability condition $[L_1, L_2]=0$ (with $L_1$ and $L_2$ from (\ref{mKP auxiltary problems})) for this equation in Lax form.

The paper is organized as follows. The first section is introduction. In second section we  reviewed basic formulae \cite{Konopelchenko&Dubrovsky2}
of $\overline\partial$-dressing method for mKP equation. Then we derived the restriction on the kernel $R_0$ of $\overline{\partial}$-problem from boundary condition $u(x,y,t)\big|_{y=0}=0$ using the so-called\, "limit of weak fields" \cite{Konopelchenko&Dubrovsky2}.
In third section we obtained general determinant formula in convenient form for the construction of exact periodical solutions of mKP-1 equation and developed the scheme of applying this formula for calculation of exact real periodical solutions. In the following forth   section we constructed by the use of proposed method new class of exact simple real nonsingular one-periodic solutions, or the class of nonlinear plane monochromatic waves. In fifths section we obtained  two-periodical solutions of mKP-1 equation without  boundary condition, then using special nonlinear superpositions of two simple one-periodical waves we calculated also  exact two-periodical solutions
  of mKP-1 equation with integrable boundary condition (\ref{BoundaryCond}). These two-periodic solutions we interpreted as eigenmodes of oscillations of the field $u(x,y,t)$   in semi-plane $y\geq 0$, the analogs of standing waves on the string with fixed endpoints.
    In conclusion we discussed the perspectives of the developed method to construction of exact periodical solutions
 for another 2+1-dimensional integrable nonlinear equations.


\section{Basic formulae of $\overline\partial$-dressing method for mKP equation, the restrictions on the kernel $R$ from reality and boundary conditions}
\label{Section_2}
\setcounter{equation}{0}
The $\overline\partial$-dressing method of Zakharov and Manakov~\cite{KonopelchenkoBook2}-\cite{Zakharov90} is founded on the use of nonlocal $\overline\partial$-problem for the wave function $\chi(\lambda,\overline\lambda)$; in the space of spectral variables $\lambda$, $\overline\lambda$ this problem is defined by kernel $R$:
\begin{equation}\label{dibar_problem}
\frac{\partial \chi (\lambda ,\overline {\lambda })}{\partial \overline {\lambda }} =
(\chi \ast R)(\lambda ,\overline{\lambda }) = \iint\limits_\mathbb{C} \chi (\lambda' ,\overline{\lambda'} )R(\lambda' ,\overline{\lambda'}  ;\lambda
,\overline{\lambda})d\lambda' \wedge d\overline\lambda'.
\end{equation}
The dependence of the kernel $R$ of $\overline\partial$-problem (\ref{dibar_problem}) and consequently of wave function $\chi$ on the physical space-time variables $x$, $y$, $t$ in the case of mKP equation (\ref{mKPv1})
has the form \cite{Konopelchenko&Dubrovsky2}, \cite{KonopelchenkoBook2}:
\begin{equation}\label{kernel}
R(\mu ,\overline \mu ;\lambda,\overline {\lambda };x,y,t)
=R_0 (\mu ,\overline \mu ;\lambda,\overline {\lambda })
e^{F(\mu; x,y,t)-F(\lambda; x,y,t)},
\end{equation}
with the function $F(\mu;x,y,t)$ given by the formula:
\begin{equation}\label{F_formula}
F(\mu;x,y,t)=i\frac{x}{\mu}+\frac{y}{\sigma\mu^2}+4i\frac{t}{\mu^3},
\end{equation}
where $\sigma=i$ for mKP-1 and $\sigma=1$ for mKP-2 equations correspondingly. The wave function $\chi$ admits the  canonical normalization:
\begin{equation}\label{normalization}
 \chi\big|_{\lambda\rightarrow\infty}\rightarrow1,
\end{equation}
such normalization follows from the special dependence $\lambda^{-1}$ of wave functions $\psi$ and $\chi$ on spectral variable $\lambda$ defined by  (\ref{mKP auxiltary problems}) and (\ref{Psi&Chi}).
From (\ref{dibar_problem}) and  (\ref{normalization}), via generalized Cauchy formula, basic singular integral equation
\begin{equation}\label{di_problem1}
\chi (\lambda,\overline {\lambda}) = 1 + \iint\limits_\mathbb{C} {\frac{d{\lambda }'\wedge
d{\overline {\lambda'}}}{2\pi i(\lambda'-\lambda)}}
\iint\limits_\mathbb{C}  \chi(\mu,\overline{\mu})
R_0(\mu ,\overline \mu ;\lambda'
,\overline {\lambda' })e^{F(\mu)-F(\lambda')}{d\mu \wedge d\overline{\mu }}
\end{equation}
of $\overline\partial$-dressing method is follows.

Reconstruction formula for the field variable $u(x,y,t)$ is given by the expression
\cite{Konopelchenko&Dubrovsky2}, \cite{KonopelchenkoBook2}:
\begin{equation}\label{reconstruct}
u=-\frac{2}{\sigma}\partial_x \ln{\chi_0},
\end{equation}
here $\chi_0$ is the zeroth order term in Taylor expansion of $\chi(\lambda,\overline\lambda;x,y,t)$ near the point $\lambda=0$:
\begin{equation}\label{Taylor}
\chi(\lambda,\overline\lambda;x,y,t)  =\chi_0(x,y,t)+\lambda\chi_1(x,y,t)+\ldots\quad.
\end{equation}
Herein and below,  for convenience,  more shorter notations such as in (\ref{di_problem1})
 \begin{equation}
F(\mu;x,y,t)\Rightarrow F(\mu), \quad \chi(\mu,\overline\mu;x,y,t)\Rightarrow\chi(\mu,\overline\mu),
\end{equation}
for $F(\mu;x,y,t), \chi(\mu,\overline\mu;x,y,t)$, etc. will be used.
Due to (\ref{di_problem1}) $\chi_0(x,y,t)$ is given by the formula:
\begin{equation}\label{chi_0}
\chi_0 (x,y,t) = 1 + \frac{2i}{\pi}\iint\limits_\mathbb{C} \frac{d\lambda_R d\lambda_I}{\lambda}
\iint\limits_\mathbb{C}  \chi(\mu,\overline{\mu})R_0(\mu ,\overline \mu ;\lambda,\overline {\lambda })e^{F(\mu)-F(\lambda)}d\mu_R  d\mu_I.
\end{equation}
The steps for construction of exact solutions via $\overline\partial$-dressing are the following. For given kernel
$R_0(\mu ,\overline \mu ;\lambda,\overline \lambda)$ one finds the solution
$\chi(\lambda,\overline\lambda)$, i.e. the wave function $\chi$, of $\overline\partial$-problem (\ref{dibar_problem}) or equivalent  integral equation (\ref{di_problem1}); then by the use of wave function $\chi$ the coefficient $\chi_0$ (\ref{chi_0}) of Taylor expansion (\ref{Taylor}) is calculated, reconstruction formula (\ref{reconstruct}) gives finally exact solution $u(x,y,t)$ of mKP-equation (\ref{mKPv1}).

Under realization of this scheme the condition of reality
 \begin{equation}\label{RealityCond}
 \overline{u(x,y,t)}=u(x,y,t),
\end{equation}
for calculation of real solutions, and integrable boundary condition
\begin{equation}\label{IntBound}
u(x,y,t)\mid_{y=0}=0,
\end{equation}
for calculation of solutions with integrable boundary, must be satisfied. Both of these conditions can be satisfied by the use of corresponding restrictions on the kernel $R$ of $\overline{\partial}$-problem (\ref{dibar_problem}); these restrictions can be obtained in so-called "limit of weak fields"\cite{Konopelchenko&Dubrovsky2}, \cite{KonopelchenkoBook2}.

Requirement of reality condition  (\ref{RealityCond}) can be satisfied by reconstruction formula  (\ref{reconstruct}) with approximate expression for (\ref{chi_0}) where in integrand (\ref{chi_0}) for $\chi_0$, in the "limit of weak fields", the first iteration $\chi(\lambda,\overline{\lambda})\approx 1$ is chosen  (see for example~\cite{Konopelchenko&Dubrovsky2}, \cite{KonopelchenkoBook2}). Such procedure leads in the case of mKP-1 to the following restriction on the kernel $R_0$:
\begin{equation}\label{RealRestr}
\mu R_0(\mu,\overline\mu;\lambda,\overline\lambda)=
\lambda\overline{R_0(\overline\lambda,\lambda;\overline\mu,\mu)}.
\end{equation}
Using the restriction (\ref{RealRestr}) in the paper \cite{Konopelchenko&Dubrovsky2} via $\overline{\partial}$-dressing several classes of exact solutions of mKP equation (\ref{mKPv1})  have been constructed: solutions with functional parameters, lumps and solitons.

The derivation of restriction on the kernel $R_0$ following from integrable boundary condition (\ref{IntBound}) we performed quite analogously. From (\ref{chi_0})  we derived via reconstruction formula (\ref{reconstruct}):
\begin{equation}\label{restrBound}
u\big|_{y=0}=-\frac{2\chi_{0x}}{\sigma\chi_0}\bigg|_{y=0}\approx
\frac{4}{\sigma\pi\chi_0}\iint\limits_\mathbb{C}\frac{d\lambda_R d\lambda_I}{\lambda}
\iint\limits_\mathbb{C}\left(\frac{1}{\mu}- \frac{1}{\lambda}\right) R_0(\mu ,\overline \mu ;\lambda,\overline {\lambda })e^{F(\mu)-F(\lambda)}\big|_{y=0}d\mu_R  d\mu_I.
\end{equation}
The expression (\ref{restrBound}) we obtained also in the limit of weak fields: for the wave function $\chi(\mu,\overline\mu)$ in integrand (\ref{restrBound}) the first iteration $\chi(\mu,\overline\mu)\approx1$ is chosen for wave function $\chi$ from equation (\ref{di_problem1}).
The phase $F(\mu)-F(\lambda)$ in exponent  (\ref{restrBound})  due to (\ref{F_formula}) has important property:
\begin{equation}\label{FF}
\left(F(-\lambda)-F(-\mu)\right)\big|_{y=0}=\left(F(\mu)-F(\lambda)\right)
\big|_{y=0},
\end{equation}
i.e. this phase does not change at $y=0$ under change of variables $\lambda\leftrightarrow-\mu$. By the change of variables $\mu\leftrightarrow-\lambda$ in integrals (\ref{restrBound}) we derived due to (\ref{FF})
\begin{equation}\label{BoundaryConditionWeakField1}
u\big|_{y=0}=-\frac{2\chi_{0x}}{\sigma\chi_0}\bigg|_{y=0}
\approx -\frac{4}{\sigma\pi\chi_0} \iint\limits_{\mathbb{C}}\frac{d\mu_{R}d\mu_{I}}{\mu}
\iint\limits_{\mathbb{C}}\left(-\frac{1}{\lambda}+\frac{1}{\mu}\right)
R_0(-\lambda,-\overline\lambda,-\mu,-\overline\mu)e^{F(\mu)-F(\lambda)}\big|_{y=0}
d\lambda_{R}d\lambda_{I}.
\end{equation}
Under the requirement
\begin{equation}\label{BoundaryConditionWeakField}
\mu R_0(\mu,\overline\mu;\lambda,\overline\lambda)=\lambda R_0(-\lambda,-\overline\lambda;-\mu,-\overline\mu)
\end{equation}
we concluded from right sides of (\ref{restrBound}) and (\ref{BoundaryConditionWeakField1}) that
\begin{equation}\label{derivBC}
u\big|_{y=0, (\ref{restrBound}) }=-u\big|_{y=0, (\ref{BoundaryConditionWeakField1}) }\Rightarrow
u\big|_{y=0}=0,
\end{equation}
the last relation means consequently the fulfillment of boundary condition (\ref{IntBound}).
For example the restriction (\ref{BoundaryConditionWeakField}) satisfies delta-form kernel $R_0$ in the form of the sum of products of delta functions:
\begin{equation}\label{RdeltaPeriod}
 R_0(\mu,\overline\mu;\lambda,\overline\lambda)=
 \sum^N_{k=1}(\lambda_k\delta(\mu-\mu_k)\delta(\lambda-\lambda_k)+
 \mu_k\delta(\mu+\lambda_k)\delta(\lambda+\mu_k)).
\end{equation}
The restrictions (\ref{RealRestr}), (\ref{BoundaryConditionWeakField}) are obtained quite analogously to each other, using the "limit of weak fields"\,: in corresponding integrands (\ref{chi_0}) and (\ref{restrBound}) approximate first iteration for wave function $\chi(\mu,\overline\mu)\approx 1$ is chosen.
In spite of non-rigorous character of derivation (\ref{derivBC}) the restriction (\ref{BoundaryConditionWeakField}) can be successfully applied for choosing appropriate kernels $R_0(\mu,\overline\mu;\lambda,\overline\lambda)$ in calculations of exact solutions of mKP-1 equation (\ref{mKPv1}) with integrable boundary condition (\ref{BoundaryCond}).

Having in mind the construction of periodical solutions $u(x,y,t)$ we used pure imaginary phases $F(\mu)-F(\lambda)$ in exponents $\exp\left(F(\mu)-F(\lambda)\right)$ of integral $\overline\partial$-equation (\ref{di_problem1}). In the case of mKP-1 equation (\ref{mKPv1}) these phases due to (\ref{F_formula}) with $\sigma=i$ have the form:
\begin{equation}\label{Phases}
F(\mu)-F(\lambda)=ix\left(\frac{1}{\mu}-\frac{1}{\lambda}\right)-iy\left(\frac{1}{\mu^2}-\frac{1}{\lambda^2}\right)+
4it\left(\frac{1}{\mu^3}-\frac{1}{\lambda^3}\right).
\end{equation}
For delta-form kernel $R_0(\mu,\overline\mu;\lambda,\overline\lambda)$
\begin{equation}\label{kernel sum}
R_0(\mu,\overline{\mu};\lambda,\overline{\lambda})
=\sum^N_{k=1} A_k\delta(\mu-\mu_k)\delta(\lambda-\lambda_k),
\end{equation}
with pure real-valued "spectral"  points $\mu_k=\overline\mu_k$, $\lambda_k=\overline\lambda_k$  in  (\ref{kernel sum})
 the phases $F(\mu)-F(\lambda)$  have pure imaginary values and corresponding exact solutions are oscillating.

Serious problem in construction of exact real periodical solutions $u$ of mKP equation (\ref{mKPv1}) is caused by simultaneous satisfaction of reality $u(x,y,t)=\overline {u(x,y,t)}$ and boundary $u(x,y,t)\mid_{y=0}=0$ conditions for calculated field. We discovered that  the restriction (\ref{RealRestr}) from reality condition, obtained in the limit of weak fields, in the case of periodical solutions does not works. We satisfied the reality condition rigorously by direct use of exact complex-valued periodical solutions in general determinant form.
In our calculations of exact real periodical solutions we followed to the steps:

A.  At first by the use of general determinant formula for exact solutions
we constructed exact complex-valued periodical (oscillating in space-time variables) solutions $u(x,y,t)$ corresponding to the delta-form kernels $R_0$ (\ref{kernel sum}) with pure real-valued spectral points $\mu_k$ and $\lambda_k$.

B. Then for complex-valued exact solutions in determinant form we derived from reality condition $u=\overline u$ the restrictions on real spectral points $\mu_k$, $\lambda_k$ and complex amplitudes $A_k\neq\overline A_k$ of delta-form kernel $R_0$ (\ref{kernel sum}).

C. Finally using obtained restrictions on parameters of complex-valued solutions we  constructed  exact real one-periodical and two-periodical solutions of mKP-1 equation in determinant form.

\section{Determinant formula for exact periodical solutions of mKP-1 equation}
\label{Section_3}
\setcounter{equation}{0}
Here we derived for given delta-kernel $R_0$ (\ref{kernel sum})
with complex amplitudes $A_k$ and real \,"spectral"\, points $\mu_k$  general determinant formula for exact periodical solutions (complex-valued in general) for mKP equation (\ref{mKPv1}). In our calculations we  followed to derivation of determinant formula for multi-soliton solutions in~\cite{Konopelchenko&Dubrovsky2}, ~\cite{KonopelchenkoBook2}.
From (\ref{di_problem1}) and (\ref{kernel sum}) we obtained the wave function $\chi(\lambda,\overline\lambda)$
\begin{equation}\label{chi(lambda1)}
\chi(\lambda,\overline\lambda)=1-\frac{2i}{\pi}\sum^N_{k=1}
\frac{A_k}{\lambda-\lambda_k}\chi(\mu_k)e^{F(\mu_k)-F(\lambda_k)}
\end{equation}
in the form of the sum of $N$ terms with simple poles at \,"spectral"\, points $\lambda_k$; such pole structure of wave function on spectral variables $\lambda$ is typical for quantum mechanics with basic Schr\"{o}dinger equation and corresponding pole structures of quantum-mechanical wave functions with poles as wave numbers, energy, momentum and, etc. Formula (\ref{chi(lambda1)}) expresses the wave function  $\chi(\lambda,\overline\lambda)$ at arbitrary values of spectral variables $\lambda, \overline\lambda$ in terms of some kind of "basis" \,or basic set of $N$ wave functions $\chi(\mu_k):=\chi(\mu_k,\overline{\mu_k})$, $(k=1,\ldots,N)$ at spectral points $\mu_k$ corresponding to the choice (\ref{kernel sum}) of the kernel $R_0$.

From (\ref{chi(lambda1)}) follows linear algebraic system of equations for the set of wave functions $\chi(\mu_k)\equiv\chi(\mu_k,\overline\mu_k)$, $(k=1,\ldots,N)$:
\begin{equation}\label{alg_syst_eq}
\sum\limits_{l=1}^{N} \tilde{A}_{kl}\chi(\mu_l)=1,\quad \tilde{A}_{kl}=\delta_{kl}+\frac{2i}{\pi}\frac{A_l e^{F(\mu_l)-F(\lambda_l)}}{\mu_k-\lambda_l},
\end{equation}
with solutions $\chi(\mu_k)$ of this system in the form:
\begin{equation}\label{chi(mu_k)}
\chi(\mu_k)=\sum\limits_{l=1}^N \tilde{A}_{kl}^{-1}.
\end{equation}
The coefficient $\chi_0(x,y,t)$ of Taylor expansion (\ref{Taylor}) due to (\ref{chi_0}), (\ref{kernel sum}) and (\ref{chi(mu_k)}) is given by expression:
\begin{equation}\label{chi0sum}
\chi_0(x,y,t)=1+\frac{2i}{\pi}\sum\limits_{k=1}^N \frac{A_k}{\lambda_k}\chi(\mu_k)e^{F(\mu_k)-F(\lambda_k)}=1+\frac{2i}{\pi}\sum\limits_{k,l=1}^N \frac{A_k}{\lambda_k}e^{F(\mu_k)-F(\lambda_k)}\tilde{A}_{kl}^{-1}.
\end{equation}
Instead of matrix $\tilde{A}_{kl}$ it is convenient to use matrix $A_{kl}$, defined by the relation:
\begin{equation}\label{relationMatrixA}
A_{kl}:=e^{F(\mu_k)}\tilde{A}_{kl}e^{-F(\mu_l)}=\delta_{kl}+
\frac{2i}{\pi}\frac{A_l e^{F(\mu_k)-F(\lambda_l)}}{\mu_k-\lambda_l},
\end{equation}
with derivatives:
\begin{equation}\label{MatrixA1}
A_{kl,x}=\frac{2A_l}{\pi}\frac{e^{F(\mu_k)-F(\lambda_l)}}{\mu_k\lambda_l}\Rightarrow A_{lk,x}=\frac{2A_k}{\pi}\frac{e^{F(\mu_l)-F(\lambda_k)}}{\mu_l\lambda_k}.
\end{equation}
For $\chi_0$ we obtained via (\ref{chi0sum})-(\ref{MatrixA1})
\begin{align}\label{chi0Matrix}
\chi_0(x,y,t)=1+i\sum\limits_{k,l=1}^N A_{lk, x}\mu_l A_{kl, x}^{-1}=1+\textrm{tr}(BA^{-1})=\ln\det(1+BA^{-1}),\nonumber
\\ B_{lk}:=i A_{lk, x}\mu_l=\frac{2iA_k}{\pi \lambda_k} e^{F(\mu_l)-F(\lambda_k)},\quad
\left(A+B\right)_{kl}=\delta_{kl}+\frac{2iA_l\mu_ke^{F(\mu_k)-
F(\lambda_l)}}{\pi\lambda_l(\mu_k-\lambda_l)}.
\end{align}
In derivation of (\ref{chi0Matrix})  we used valid for rank-1 matrix $B$ the following identity:
\begin{equation}\label{chi0Matrix1}
1+\textrm{tr}(BA^{-1})=\ln\det(1+BA^{-1}).
\end{equation}
Finally from  (\ref{chi0Matrix}) by reconstruction formula (\ref{reconstruct}) follows general determinant formula for exact solutions of mKP equation $u(x,y,t)$  corresponding to delta-form kernel (\ref{kernel sum}):
\begin{equation}\label{reconstructMatrix}
u=-\frac{2}{\sigma}\frac{\partial}{\partial x}\ln\det(1+BA^{-1})=-\frac{2}{\sigma}\frac{\partial}{\partial x}\ln\det\frac{A+B}{A},
\end{equation}
this expression is valid for both types mKP-1 and mKP-2 equations. In general this formula gives complex-valued exact solutions.

For the investigation of restrictions on parameters $A_k$ and $\mu_k$, $\lambda_k$ of the kernel $R_0$ following from reality condition $u=\overline u$ it is convenient to transform matrices $A+B$ and $A$ from (\ref{relationMatrixA})-(\ref{reconstructMatrix}) to some equivalent and more symmetrical form. Using definitions (\ref{chi0Matrix}) we introduced matrix $N$:
\begin{equation}\label{Matrix A+B}
\left(A+B\right)_{kl}=\delta_{kl}+\frac{2iA_l\mu_ke^{F(\mu_k)-
F(\lambda_l)}}{\pi\lambda_l(\mu_k-\lambda_l)}:=e^{\frac{F(\mu_k)}{2}}N_{kl}
\frac{2A_l}{\pi\lambda_l}e^{-\frac{F(\lambda_l)}{2}},
\end{equation}
where
\begin{equation}\label{MatrixN}
N_{kl}:=\delta_{kl}\frac{\pi\lambda_k}{2A_k}e^{-\frac{\Delta F(\mu_k,\lambda_l)}{2}}+\frac{i\mu_k}{\mu_k-\lambda_l}e^{\frac{\Delta F(\mu_k,\lambda_l)}{2}}.
\end{equation}
Analogously instead of  matrix $A$ we introduced matrix $D$:
\begin{equation}\label{MatrixA2}
A_{kl}=\delta_{kl}+\frac{2iA_l}{\pi(\mu_k-\lambda_l)}e^{F(\mu_k)-F(\lambda_l)}:=e^{\frac{F(\mu_k)}{2}}D_{kl}
\frac{2A_l}{\pi\lambda_l}e^{-\frac{F(\lambda_l)}{2}},
\end{equation}
where
\begin{equation}\label{MatrixD}
D_{kl}:=\delta_{kl}\frac{\pi\lambda_k}{2A_k}e^{-\frac{\Delta F(\mu_k,\lambda_l)}{2}}+\frac{i\lambda_l}{\mu_k-\lambda_l}e^{\frac{\Delta F(\mu_k,\lambda_l)}{2}},
\end{equation}
here in (\ref{MatrixN}),  (\ref{MatrixD}) we used the notations:
\begin{equation}\label{DeltaF}
\Delta F(\mu_k, \lambda_l):=F(\mu_k)-F(\lambda_l)= ix\left(\frac{1}{\mu_k}-\frac{1}{\lambda_l}\right)-iy\left(\frac{1}{\mu^2_k}-\frac{1}{\lambda^2_l}\right)+
4it\left(\frac{1}{\mu^3_k}-\frac{1}{\lambda^3_l}\right).
\end{equation}
Evidently for real spectral points $\mu_k$ and $\lambda_k$ of delta-form kernel $R_0$ (\ref{kernel sum}) the quantities $\Delta F(\mu_k, \lambda_l)$ satisfy to relations:
\begin{equation}\label{imaginaryCond}
\overline{\Delta F(\mu_k, \lambda_l)}=\Delta F(\lambda_l, \mu_k).
\end{equation}

The reconstruction formula (\ref{reconstructMatrix}) for mKP-1 with $\sigma=i$, due to (\ref{Matrix A+B})-(\ref{MatrixD}), transforms to the following equivalent form in terms of matrices $N$ and $D$:
\begin{equation}\label{Reconstruct1}
u=-\frac{2}{\sigma}\frac{\partial}{\partial x}\ln\det\frac{N}{D}=2i\frac{\partial}{\partial x}\ln\det\frac{N}{D}.
\end{equation}
The requirement of reality of $u$ field $u=\overline u$, due to (\ref{MatrixN}), (\ref{MatrixD}), (\ref{imaginaryCond}) and (\ref{Reconstruct1}), in the case of mKP-1 equation leads to the condition:
\begin{align}\label{realityCond}
\det{N}=\det\left(\delta_{kl}\frac{\pi\lambda_k}{2A_k}e^{-\frac{\Delta F(\mu_k,\lambda_l)}{2}}+\frac{i\mu_k}{\mu_k-\lambda_l}e^{\frac{\Delta F(\mu_k,\lambda_l)}{2}}\right)=\overline{\det D}=\nonumber \\
=\det\left(\delta_{kl}\frac{\pi\lambda_k}{2\overline A_k}
e^{-\frac{\Delta F(\lambda_l,\mu_k)}{2}}-\frac{i\lambda_l}{\mu_k-\lambda_l}e^{\frac{\Delta F(\lambda_l,\mu_k)}{2}}\right).
\end{align}
If the condition (\ref{realityCond}) is satisfied, the reconstruction formula (\ref{Reconstruct1}) gives for exact real solutions of mKP-1 equation the following expression:
\begin{equation}\label{reconstruct1}
u(x,y,t)=2i\frac{\partial}{\partial x}\left(-2i \arg(\det D)\right)=4\frac{\partial}{\partial x} \arctan\frac{\det_I D}{\det_R D},
\end{equation}
here we used real $\det_R D$ and imaginary $\det_I D$ parts of $\det D=\det_R D+i\det_I D$.

In the following sections 4 and 5 we presented new classes of  exact real one-periodical and two-periodical solutions of mKP-1 equation.

\section{New class of exact one-periodic solutions of mKP-1 equation, nonlinear plane monochromatic waves}
\label{Section_4}
\begin{figure}[h]
\begin{center}
\includegraphics[width=0.50\textwidth,keepaspectratio]{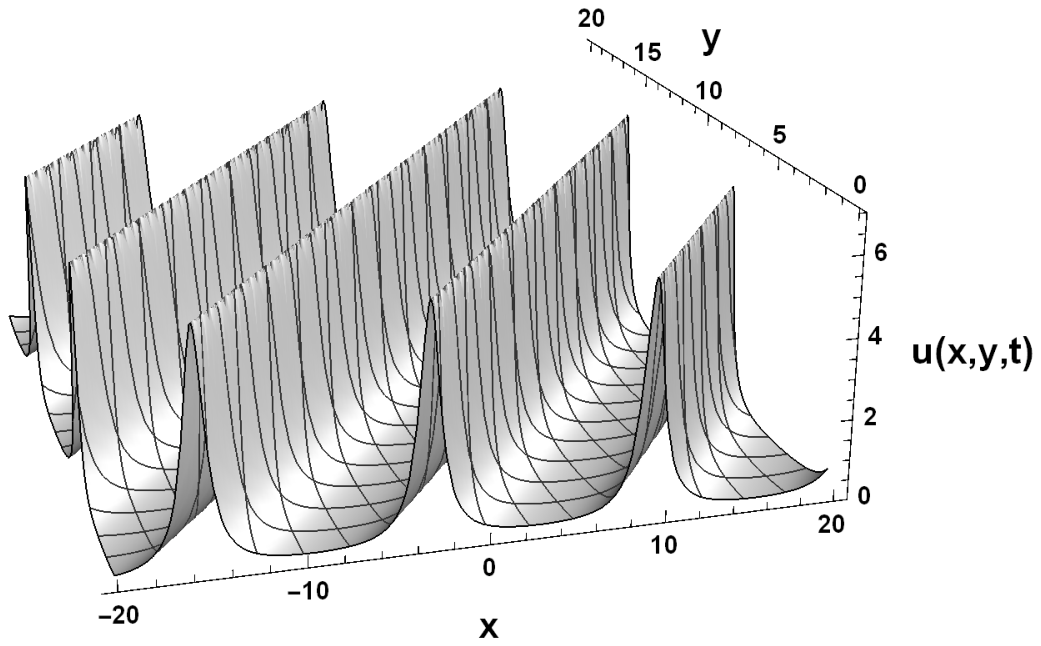}
\parbox[t]{1\textwidth}{\caption{Periodic solution $u$ (\ref{SolutionExample})  with parameters $\lambda_{0}=1$, $\mu_{0}=2$.}\label{OnePeriodic(RealSpectral)mKP1}}
\end{center}
\end{figure}

\setcounter{equation}{0}
For the calculation of simplest periodic solution of mKP-1 equation we considered simple delta-form kernel $R_0$
\begin{equation}\label{kernelExample}
R_0(\mu,\overline\mu;\lambda,\overline\lambda)=A_0\delta(\mu-\mu_0)
\delta(\lambda-\lambda_0)
\end{equation}
in $\overline\partial$-dressing integral equation (\ref{di_problem1}) with complex amplitude $A_0$ and real\, "spectral"\, points $\mu_0=\overline\mu_0$, $\lambda_0=\overline\lambda_0$. We derived for $\det N$ and $\det D$ due to (\ref{Matrix A+B})-(\ref{MatrixD})  the following expressions:
\begin{align}
\det{N}=\left(\frac{\pi\lambda_0}{2A_0}e^{-\frac{\Delta F(\mu_0,\lambda_0)}{2}}+\frac{i\mu_0}{\mu_0-\lambda_0}e^{\frac{\Delta F(\mu_0,\lambda_0)}{2}}\right),\nonumber \\
\det{D}=\left(\frac{\pi\lambda_0}{2A_0}e^{-\frac{\Delta F(\mu_0,\lambda_0)}{2}}+\frac{i\lambda_0}{\mu_0-\lambda_0}e^{\frac{\Delta F(\mu_0,\lambda_0)}{2}}\right),
\end{align}
here accoding (\ref{DeltaF}):
\begin{equation}
\Delta F(\mu_0, \lambda_0)=ix\left(\frac{1}{\mu_0}-\frac{1}{\lambda_0}\right)-iy\left(\frac{1}{\mu^2_0}-\frac{1}{\lambda^2_0}\right)+
4it\left(\frac{1}{\mu^3_0}-\frac{1}{\lambda^3_0}\right):=i\Delta\tilde\Phi,
\end{equation}
here $\overline{\Delta\tilde\Phi}=\Delta\tilde\Phi$. The requirement of reality of $u$ field $u=\overline u$ gives in considered case:
\begin{align}\label{realityCondExmple}
\det{N}=\left(\frac{\pi\lambda_0}{2A_0}e^{-\frac{i\Delta\tilde\Phi}{2}}+
\frac{i\mu_0}{\mu_0-\lambda_0}e^{\frac{i\Delta\tilde\Phi}{2}}\right)=
\overline{\det{D}}= \nonumber\\
=\left(\frac{\pi\lambda_0}{2\overline A_0}e^{\frac{i\Delta\tilde\Phi}{2}}-\frac{i\lambda_0}{\mu_0-\lambda_0}
e^{-\frac{i\Delta\tilde\Phi}{2}}\right).
\end{align}
From (\ref{realityCondExmple}) we concluded:
\begin{equation}
\frac{\pi\lambda_0}{2A_0}=-\frac{i\lambda_0}{\mu_0-\lambda_0},\quad
\frac{\pi\lambda_0}{2\overline A_0}=\frac{i\mu_0}{\mu_0-\lambda_0},
\end{equation}
and consequently
\begin{equation}\label{AmplitudeExample}
|A_0|=\frac{\pi}{2}\sqrt{\frac{\lambda_0}{\mu_0}}(\mu_0-\lambda_0),\quad A_0=|A_0|e^{i\alpha},
\end{equation}
with  real constant $\alpha=\arg A$. Inserting (\ref{AmplitudeExample}) into (\ref{realityCondExmple}) we obtained:
\begin{equation}\label{DetD}
\det D=\left(\frac{\sqrt{\mu_0\lambda_0}}{\mu_0-\lambda_0}e^{-i\frac{\Delta\tilde\Phi+\alpha}{2}}
+\frac{i\lambda_0}{\mu_0-\lambda_0}e^{i\frac{\Delta\tilde\Phi+\alpha}{2}}
\right)e^{-i\frac{\alpha}{2}},
\end{equation}
and from (\ref{DetD})
\begin{equation}\label{TgArgDet}
\arg\det D=\arctan\left(\frac{\sqrt{\frac{\lambda_0}{\mu_0}}\cos\frac{\Delta\Phi}{2}-\sin\frac{\Delta\Phi}{2}}
{\cos\frac{\Delta\Phi}{2}-\sqrt{\frac{\lambda_0}{\mu_0}}\sin\frac{\Delta\Phi}{2}}\right)-\frac{\alpha}{2},
\quad \Delta\Phi:=\Delta\tilde\Phi+\alpha.
\end{equation}
Finally using the reconstruction formula (\ref{reconstruct1}) with (\ref{TgArgDet}) after simple calculations we derived exact periodic solution of mKP-1 equation:
\begin{equation}\label{SolutionExample}
u=4\frac{\partial}{\partial x}\left(\arg\det D\right)= 4 \frac{\frac{(\lambda_0-\mu_0)^2}{2\lambda_0\mu^2_0}}
{1+\frac{\lambda_0}{\mu_0}-2\sqrt{\frac{\lambda_0}{\mu_0}}\sin\Delta\Phi}=
\frac{4\sinh^2\Xi}{\sqrt{\lambda_0\mu_0}\left(\cosh\Xi-
\sin\Delta\Phi\right)},
\end{equation}
here $e^{\Xi}:=\sqrt{\frac{\lambda_0}{\mu_0}}$, \quad $\Delta\Phi=x\left(\frac{1}{\mu_0}-\frac{1}{\lambda_0}\right)-y\left(\frac{1}{\mu^2_0}-\frac{1}{\lambda^2_0}\right)+
4t\left(\frac{1}{\mu^3_0}-\frac{1}{\lambda^3_0}\right)+\alpha$. The graph of this solution is shown on figure (\ref{OnePeriodic(RealSpectral)mKP1}). Rewriting phase $\Delta\Phi$ in convenient form:
\begin{equation}\label{PhaseExample}
\Delta\Phi=k_x x+k_y y-\omega t+\alpha,
\end{equation}
we concluded that field $u$ propagates harmonically in plane $(x,y)$ in direction of wave vector $\vec{k}$ and oscillates in time with frequency $\omega$ defined by the expressions:
\begin{equation}\label{Direction}
\vec{k}=(k_x,k_y)=\left(\frac{1}{\mu_0}-\frac{1}{\lambda_0},-\frac{1}{\mu^2_0}+
\frac{1}{\lambda^2_0}\right), \quad
\omega= 4 \left|\frac{1}{\mu^3_0}-\frac{1}{\lambda^3_0}\right|.
\end{equation}
The velocity of this periodical nonlinear wave is given by expression:
\begin{equation}\label{velocity}
V:=\frac{|\omega|}{|\vec{k}|}=\frac{4 \left|\frac{1}{\mu^3_0}-\frac{1}{\lambda^3_0}\right|}{\left|\frac{1}{\mu_0}-\frac{1}{\lambda_0}\right|
\sqrt{1+\left(\frac{1}{\mu_0}+\frac{1}{\lambda_0}\right)^2}}=
\frac{4 \left(\frac{1}{\mu^2_0}+\frac{1}{\lambda^2_0}+\frac{1}{\mu_0\lambda_0}\right)}
{\sqrt{1+\left(\frac{1}{\mu_0}+\frac{1}{\lambda_0}\right)^2}}.
\end{equation}
Exact one-periodic solution (\ref{SolutionExample}) due to $\cosh\Xi>1$ is nonsingular. This solution is representative of new class of line-periodic harmonic solutions or nonlinear plane monochromatic waves of mKP-1 equation.
This new class we named as  analog of the class of plane monochromatic waves for linear wave equations.

More general kernel $R_0$ (\ref{kernel sum}) with $N$ delta-functional terms leads to nonlinear superposition of simple nonlinear plane monochromatic waves. For kernel (\ref{kernel sum}) with two terms in the sum we constructed in next section new classes of exact real two-periodical solutions of mKP-1 equation.

\section{New classes of exact two-periodic solutions of mKP-1 equation}
\label{Section_5}
\setcounter{equation}{0}
Two-periodic solution of mKP-1 equation corresponds to delta-form kernel $R_0$ with two  terms of the type (\ref{kernelExample}):
\begin{equation}\label{kernel sumExample}
R_0(\mu,\overline{\mu};\lambda,\overline{\lambda})=A_1\delta(\mu-\mu_0)\delta(\lambda-\lambda_0)+
A_2\delta(\mu+\lambda_0)\delta(\lambda+\mu_0)
\end{equation}
with complex amplitudes $A_1$, $A_2$ and real spectral points $\mu_0=\overline{\mu_0}$, $\lambda_0=\overline{\lambda_0}$. Matrices $N$ and $D$ due to (\ref{MatrixN}), (\ref{MatrixD}) have the forms:
\begin{equation}\label{MatrixN_Example}
N=\left(
  \begin{array}{cc}
    \frac{\pi\lambda_0}{2A_1}e^{-\frac{\Delta F(\mu_0,\lambda_0)}{2}} +\frac{i\mu_0}{\mu_0-\lambda_0}e^{\frac{\Delta F(\mu_0,\lambda_0)}{2}}& \frac{i\mu_0}{\mu_0+\mu_0}e^{\frac{\Delta F(\mu_0,-\mu_0)}{2}}\\
   \frac{-i\lambda_0}{-\lambda_0-\lambda_0}e^{\frac{\Delta F(-\lambda_0,\lambda_0)}{2}} &  -\frac{\pi\mu_0}{2A_2}e^{-\frac{\Delta F(-\lambda_0,-\mu_0)}{2}} -\frac{i\lambda_0}{-\lambda_0+\mu_0}e^{\frac{\Delta F(-\lambda_0,-\mu_0)}{2}} \\
  \end{array}
\right),
\end{equation}
\begin{equation}\label{MatrixD_Example}
D=\left(
  \begin{array}{cc}
    \frac{\pi\lambda_0}{2A_1}e^{-\frac{\Delta F(\mu_0,\lambda_0)}{2}} +\frac{i\lambda_0}{\mu_0-\lambda_0}e^{\frac{\Delta F(\mu_0,\lambda_0)}{2}}& \frac{-i\mu_0}{\mu_0+\mu_0}e^{\frac{\Delta F(\mu_0,-\mu_0)}{2}}\\
   \frac{i\lambda_0}{-\lambda_0-\lambda_0}e^{\frac{\Delta F(-\lambda_0,\lambda_0)}{2}} &  -\frac{\pi\mu_0}{2A_2}e^{-\frac{\Delta F(-\lambda_0,-\mu_0)}{2}} -\frac{i\mu_0}{-\lambda_0+\mu_0}e^{\frac{\Delta F(-\lambda_0,-\mu_0)}{2}} \\
  \end{array}
\right).
\end{equation}
In (\ref{MatrixN_Example}) and (\ref{MatrixD_Example}) the phases  in exponents  due to  (\ref{DeltaF}) are the following:
\begin{align}\label{DeltaFExample}
&\Delta F(\mu_0, \lambda_0)= ix\left(\frac{1}{\mu_0}-\frac{1}{\lambda_0}\right)-
iy\left(\frac{1}{\mu^2_0}-\frac{1}{\lambda^2_0}\right)+
4it\left(\frac{1}{\mu^3_0}-\frac{1}{\lambda^3_0}\right)
:= i\left(\Phi(x,t)-\Theta(y)\right), \nonumber\\
&\Delta F(-\lambda_0, -\mu_0)=i\left(\Phi(x,t)+\Theta(y)\right),\quad \Delta F(\mu_0, -\mu_0)+\Delta F(-\lambda_0, \lambda_0)=2i\Phi(x,t), \nonumber\\
&\Delta F(-\lambda_0, \lambda_0)=-2i\frac{x}{\lambda_0}-8it\frac{1}{\lambda^3_0},\quad
\Delta F(\mu_0, -\mu_0)=2i\frac{x}{\mu_0}+8it\frac{1}{\mu^3_0},
\nonumber\\
&\Phi(x,t):= x\left(\frac{1}{\mu_0}-\frac{1}{\lambda_0}\right)
+ 4t\left(\frac{1}{\mu^3_0}-\frac{1}{\lambda^3_0}\right),\quad
\Theta(y):=y\left(\frac{1}{\mu^2_0}-\frac{1}{\lambda^2_0}\right).
\end{align}
Rewriting $N$ and $\overline D$ in notations of (\ref{DeltaFExample}) we obtained:
\begin{equation}\label{MatrixN_Example1}
N=\left(
  \begin{array}{cc}
    \frac{\pi\lambda_0}{2A_1}e^{-i\frac{\Phi-\Theta}{2}} +\frac{i\mu_0}{\mu_0-\lambda_0}e^{i\frac{\Phi-\Theta}{2}}& \frac{i}{2}e^{i\frac{x}{\mu_0}+4i\frac{t}{\mu^3_0}}\\
   \frac{i}{2}e^{-i\frac{x}{\lambda_0}-4i\frac{t}{\lambda^3_0}}&  -\frac{\pi\mu_0}{2A_2}e^{-i\frac{\Phi+\Theta}{2}} +\frac{i\lambda_0}{\lambda_0-\mu_0}e^{i\frac{\Phi+\Theta}{2}} \\
  \end{array}
\right),
\end{equation}
\begin{equation}\label{MatrixD_Example1}
\overline D=\left(
  \begin{array}{cc}
    \frac{\pi\lambda_0}{2\overline A_1}e^{i\frac{\Phi-\Theta}{2}} -\frac{i\lambda_0}{\mu_0-\lambda_0}e^{-i\frac{\Phi-\Theta}{2}}& \frac{i}{2}e^{-i\frac{x}{\mu_0}-4i\frac{t}{\mu^3_0}}\\
   \frac{i}{2}e^{i\frac{x}{\lambda_0}+4i\frac{t}{\lambda^3_0}} &  -\frac{\pi\mu_0}{2\overline A_2}e^{i\frac{\Phi+\Theta}{2}} +\frac{i\mu_0}{\mu_0-\lambda_0}e^{-i\frac{\Phi+\Theta}{2}} \\
  \end{array}
\right).
\end{equation}
Requirement of reality of $u$ field $\overline u=u$, which due to (\ref{Reconstruct1}) can be expressed in the form (\ref{realityCond})
with matrices $N$ and $\overline D$ given by (\ref{MatrixN_Example1}) and (\ref{MatrixD_Example1}), leads to the restrictions on complex amplitudes $A_1$, $A_2$ and spectral points $\mu_0=\overline{\mu_0}$, $\lambda_0=\overline{\lambda_0}$. Equating the coefficients at the exponents $e^{\pm i\Phi}$, $e^{\pm i\Theta}$ in $\det N=\overline {\det D}$ we derived the relations:
\begin{align}\label{restrictionsExample}
  (e^{i\Phi}):\;-\frac{\pi^2\lambda_0\mu_0}{4\overline{A_1}\cdot\overline{A_2}} = \frac{(\mu_0+\lambda_0)^2}{4(\mu_0-\lambda_0)^2};\quad
  (e^{-i\Phi}):\;-\frac{\pi^2\lambda_0\mu_0}{4A_1A_2} = \frac{(\mu_0+\lambda_0)^2}{4(\mu_0-\lambda_0)^2};\nonumber \\
  (e^{i\Theta}):\;\frac{i\pi\lambda_0\mu_0}{2\overline{A_2}(\mu_0-\lambda_0)} = -\frac{i\pi\lambda_0^2}{2(\mu_0-\lambda_0) A_1}; \quad
  (e^{-i\Theta}):\;\frac{i\pi\lambda_0\mu_0}{2\overline{A_1}(\mu_0-\lambda_0)} = -\frac{i\pi\mu_0^2}{2(\mu_0-\lambda_0) A_2}.
\end{align}
The restrictions (\ref{restrictionsExample}) on $A_1$, $A_2$ and $\mu_0$, $\lambda_0$ can be satisfied for the choice
\begin{equation}\label{ConditParam}
\overline A_2=-A_1\frac{\mu_0}{\lambda_0},\quad A_1=|A_1|e^{i\alpha}, \quad |A_1|=\frac{\pi\lambda_0(\mu_0-\lambda_0)}{(\mu_0+\lambda_0)},
\end{equation}
here $\alpha$ is arbitrary real constant:
\begin{equation}\label{restrictionsExample1}
A_1=|A_1|e^{i\alpha}=\frac{\pi\lambda_0(\mu_0-\lambda_0)}{(\mu_0+\lambda_0)}e^{i\alpha},\quad
A_2=-\overline A_1\frac{\mu_0}{\lambda_0}=-\frac{\pi\mu_0(\mu_0-\lambda_0)}
{(\mu_0+\lambda_0)}e^{-i\alpha}.
\end{equation}
From (\ref{MatrixD_Example1}) and (\ref{restrictionsExample1}) follows the expression for $\det D$:
\begin{equation}\label{DetDExample}
\det D=\frac{(\mu_0+\lambda_0)^2}{2(\mu_0-\lambda_0)^2}\cos\Phi(x,t)+\frac{(\mu_0+\lambda_0)^2}{2(\mu_0-\lambda_0)^2}
\sin\left(\Theta(y)-\alpha\right)-\frac{i(\mu_0+\lambda_0)}
{2(\mu_0-\lambda_0)}\cos\left(\Theta(y)-\alpha\right).
\end{equation}
So the requirement of reality $u=\overline u$, due to $\det N=\overline{\det D}$ and relations (\ref{restrictionsExample1}), is fulfilled and
$\tan(\arg\det D)$ due to (\ref{DetDExample}) has the form:
\begin{equation}\label{TgArgDetEx}
\tan(\arg\det D)=-\frac{\frac{(\mu_0-\lambda_0)}{(\mu_0+\lambda_0)}\cos
\left(\Theta-\alpha\right)}{\cos\Phi+\sin\left(\Theta-\alpha\right)},
\end{equation}
here due to (\ref{DeltaFExample}) and (\ref{restrictionsExample1})
\begin{equation}\label{phases}
\alpha=\arg A_1,\quad \Phi(x,t)=x\left(\frac{1}{\mu_0}-\frac{1}{\lambda_0}\right)+4t\left(\frac{1}{\mu^3_0}-\frac{1}{\lambda^3_0}\right),\quad
\Theta(y)=y\left(\frac{1}{\mu^2_0}-\frac{1}{\lambda^2_0}\right).
\end{equation}

Via (\ref{TgArgDetEx}) using  reconstruction formula (\ref{reconstruct1}) for $u$ we obtained the following two-periodic real solution of mKP-1 equation:
\begin{align}\label{TwoPeriodWithoutBound}
u(x,y,t)=\frac{\frac{4(\mu_0-\lambda_0)^2}{\lambda_0\mu_0(\mu_0+\lambda_0)}\sin\Phi(x,t)\cos\left(\Theta(y)-\alpha\right)}
{\left(\cos\Phi(x,t)+\sin\left(\Theta-\alpha\right)\right)^2+\left(\frac{\mu_0-\lambda_0}{\mu_0+\lambda_0}\right)^2
\cos^2\left(\Theta(y)-\alpha\right)}=\nonumber \\
\frac{8\sinh^2\Xi}{\sqrt{\mu_0\lambda_0}\cosh\Xi}\frac{\sin\Phi(x,t)\cos\left(\Theta(y)-\alpha\right)}{\left(\cos\Phi(x,t)+\sin
\left(\Theta(y)-\alpha\right)\right)^2+
\tanh^2\Xi\cos^2\left(\Theta(y)-\alpha\right)},
\end{align}
here $e^\Xi=\sqrt{\frac{\lambda_0}{\mu_0}}$.
Solution (\ref{TwoPeriodWithoutBound}) corresponds to the delta-form kernel $R_0$ of the form (\ref{kernel sumExample}) with two  terms of the type (\ref{kernelExample}); this solution is representative of new class  of exact two-periodical solutions.  Two-periodical solutions of the type (\ref{TwoPeriodWithoutBound}) do not satisfy to integrable boundary condition (\ref{BoundaryCond}).

In order to construct exact solutions with integrable boundary (\ref{BoundaryCond}) we  taken in to account additional restriction on kernel $R_0$ (\ref{BoundaryConditionWeakField}), obtained in the limit of weak fields. Such restriction is satisfied for the kernels of the type (\ref{RdeltaPeriod}), i. e. to the kernel (\ref{kernel sumExample}) with the following relation between amplitudes $A_1$ and $A_2$:
\begin{equation}\label{CondFromBound}
 A_2=A_1\frac{\mu_0}{\lambda_0}.
\end{equation}
From the relations (\ref{ConditParam})
and (\ref{CondFromBound}) between amplitudes $A_1$ and $A_2$ taken together we derived:
\begin{equation}\label{FinRelAmpl}
 A_2=-\overline{A_1}\frac{\mu_0}{\lambda_0}=A_1\frac{\mu_0}{\lambda_0},
\end{equation}
and consequently
\begin{equation}\label{FinRelAmpl1}
 \alpha=\arg A_1=\frac{\pi}{2}.
\end{equation}
\begin{figure}[h]
\begin{center}
\includegraphics[width=0.7\textwidth,keepaspectratio]{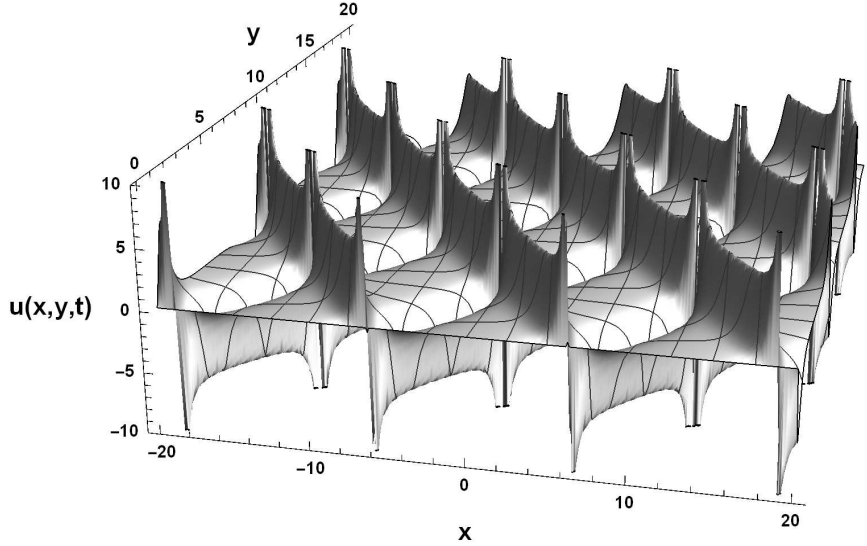}
\parbox[t]{1\textwidth}{\caption{Two-periodic solution $u$ (\ref{TwoPeriodicSolution})  with parameters: $\lambda_{0}=1$, $\mu_{0}=2$.}\label{Periodic(RealSpectral)mKP1}}
\end{center}
\end{figure}
So due to (\ref{FinRelAmpl1}) under the additional choice of constant phase $\arg A_1=\alpha=\frac{\pi}{2}$,
we derived the exact real two-periodic solution (\ref{TwoPeriodWithoutBound}) of mKP-1 equation with integrable boundary (\ref{BoundaryCond}):
\begin{equation}\label{TwoPeriodicSolution}
u(x,y,t)=\frac{8\tanh\Xi\sinh\Xi}{\sqrt{\mu_0\lambda_0}}\frac{\sin\Phi(x,t)\sin\Theta(y)}
{\left(\cos\Phi(x,t)-\cos\Theta(y)\right)^2+\tanh^2\Xi\sin^2\Theta(y)},
\end{equation}
here $e^\Xi=\sqrt{\frac{\lambda_0}{\mu_0}}$. The graph of this solution is shown on figure (\ref{Periodic(RealSpectral)mKP1}). Obtained solution (\ref{TwoPeriodicSolution}) by construction satisfies to boundary condition (\ref{BoundaryCond}). This solution is two-periodical on phases $\Phi(x,t)$  and $\Theta(y)$ (\ref{phases})
and represents nonlinear harmonic wave: static-periodical in direction of axis $y$ and propagating harmonically along axis $x$ with velocity:
\begin{equation}\label{1v}
V_x=-4\frac{\frac{1}{\mu^3_0}-\frac{1}{\lambda^3_0}}
{\frac{1}{\mu_0}-\frac{1}{\lambda_0}}=
-4\left(\frac{1}{\mu^2_0}+\frac{1}{\lambda^2_0}+\frac{1}{\mu_0\lambda_0}\right).
\end{equation}
This solution has point singularities defined by equations:
\begin{equation}\label{Singul}
\sin\Theta(y)=0, \quad \cos\Phi(x,t)-\cos\Theta(y)=0.
\end{equation}
It follows from equations (\ref{Singul}) that singularities of exact solution (\ref{TwoPeriodicSolution}) at fixed moment of time $t$ are arranged periodically on the plane $(x,y)$ with positions at the points:
\begin{equation}
\Phi(x,t)=x\left(\frac{1}{\mu_0}-\frac{1}{\lambda_0}\right)
+ 4t\left(\frac{1}{\mu^3_0}-\frac{1}{\lambda^3_0}\right)=n\pi,\quad
\Theta(y)=y\left(\frac{1}{\mu^2_0}-\frac{1}{\lambda^2_0}\right)=m\pi,
\end{equation}
here $n,m=0,1,\ldots$. The solution (\ref{TwoPeriodicSolution}) is representative of new class  of exact two-periodical solutions of mKP-1 equation with  integrable boundary condition $u(x,y,t)\mid_{y=0}=0$ (\ref{BoundaryCond}).

\section{Conclusions}
\label{Section_7}
\setcounter{equation}{0}
We described new classes of exact periodical solutions of mKP-1 equation and developed general scheme for their calculations in the framework of $\overline\partial$-dressing method. We satisfied the reality condition for solutions $u$ exactly, imposing the requirement of reality $u=\overline u$ on exact complex-valued solutions in general determinant form; by this way we obtained certain  restrictions on parameters of solutions, i.e. on amplitudes $A_k$ and spectral points $\mu_k$, $\lambda_k$ of delta-form kernel $R_0$  of $\overline\partial$-problem.

We constructed new class of exact one-periodic nonsingular solutions, the class of nonlinear plane monochromatic waves. These solutions can be named line-periodic, in close analogy with line-soliton solutions of two-dimensional integrable nonlinear equations.

We calculated two new classes of real exact two-periodic solutions: the class of solutions  without any boundary conditions and the specialized class of solutions with integrable boundary (\ref{BoundaryCond}); both types of solutions have periodically arranged point singularities.  The imposition on the field $u$ of boundary condition (\ref{BoundaryCond}) leads to formation of certain eigenmodes of oscillations of the field $u(x,y,t)$  on semi-plane $y\geq 0$  with fixed maxima and minima along $y$-axis, this field propagates harmonically with certain velocity along $x$-axis.

We demonstrated effectiveness of $\overline\partial$-dressing method in calculations of periodical solutions of several types for mKP-1 equation.
The developed in present paper procedure for calculations via $\overline\partial$-dressing of new classes of exact real one-periodic and two-periodic solutions
solutions can be effectively applied to all other integrable (2+1)-dimensional nonlinear equations, the
program of such researches now in progress and corresponding results will be published elsewhere.



\end{document}